\documentclass[a4paper,11pt]{article}
\usepackage{axodraw}
\usepackage{graphicx}
\def\hyg{{~_2F_1}}
\def\eps{\varepsilon}

\def\eir{\varepsilon_{IR}}
\def\hyg{{_2F_1}}
\def\s{{\tilde{s}}}
\def\t{{\tilde{t}}}
\def\u{{\tilde{u}}}

\def\m2{{\tilde{m_2^2}}}
\def\sm3{{\tilde{m_3^2}}}

\begin{document}
\begin{flushright}
KEK-CP-159
\end{flushright}
\vskip 2cm
\begin{center}
{\Large
Numerical Contour Integration for Loop Integrals
}
\end{center}
\vskip 1cm
\begin{center}
{\large Y.~Kurihara and T.~Kaneko
}\\
\vskip 0.5cm
{\it High Energy Accelerator Research Organization,\\
Oho 1-1, Tsukuba, Ibaraki 305-0801, Japan}\\
\end{center}
\vskip 3cm

\begin{abstract}
A fully numerical method to calculate loop integrals,
a numerical contour-integration method, is proposed.
Loop integrals can be interpreted as a contour integral
in a complex plane for an integrand with multi-poles
in the plane. Stable and efficient numerical integrations
an along appropriate contour can be performed for tensor integrals
as well as for scalar ones appearing in loop calculations of
the standard model.
Examples of 3- and 4-point diagrams in 1-loop integrals
and 2- and 3-point diagrams in 2-loop integrals 
with arbitrary masses are shown.

Moreover it is shown that 
numerical evaluations of the Hypergeometric function, which 
often appears in the loop integrals, can be performed using the numerical
contour-integration method.
\end{abstract}
\newpage

\section{Introduction}
In future collider experiments such as at the LHC and ILC,
the standard model will be checked with very high precision and a signal 
of new physics will be searched for through a tiny difference between
experimental measurements and theoretical predictions.
The theoretical uncertainty must be, at least, one order of magnitude
smaller than the experimental one. The theoretical prediction is obtained
based on the perturbative calculation of the quantum field theories. 
The theoretical calculation of a higher order correction 
with one- or two-loops must be performed 
in order to meet with experimental requirements.
A loop integration is one of the critical issues of 
a computation of these higher order corrections.
Basically there are two methods to perform the loop integration, 
an analytical method 
and a numerical one. Though the analytical method gives, in principle, fast and
stable results, for many cases it is hard to give a compact expression
for the one-loop integral
and it is very difficult to perform two-loop computations
for the standard model with multiple energy scales.

Numerical methods have also been investigated to
perform loop integrals. The symmetrical
sampling method was proposed in \cite{numeri1}
by Oyanagi {\it et al.} 
in 1988 and has been investigated through
a series of papers\cite{numeri2,numeri3,numeri4}.
Another numerical method, the hybrid method\cite{2loop1}, has been proposed.
It is to perform a part of the multi-dimensional integrations
analytically up to the integrand being a logarithmic form.
A remaining integration was performed numerically by Monte Carlo
method. Both methods are tuned to calculate up to two-loop/three-point 
functions with arbitrary masses\cite{2loop2}.

Recently another numerical 
methods using the Sato-Bernstein-Tkachov (SBT) relation
\footnote{This relation was introduced at first by Sato\cite{sato}
relating to prehomogeneous vector spaces and was generally proved
by Bernstein\cite{bernstein}. An application of this relation 
to the loop integral was proposed by Tkachov\cite{tkachov}.}
is proposed and utilized 
intensively\cite{passarino1,passarino2,weber}. 
A recent review of the method using the SBT relation
can be found in \cite{uccirati}.

Yet another numerical method has been proposed by de Doncker\cite{deDoncker},
an `$\eps$-algorithm'. This method takes 
the infinitesimal imaginary parameter appearing in a denominator
of the loop integral at finite values, and takes its extrapolation to zero.
It is demonstrated that 
this method can give precise results for a non-scalar one-loop/three-point 
digram\cite{deDoncker}.

We propose a new method of loop integrations
such as a `numerical contour-integration (NCI) method' in this report. 
The loop integral can be interpreted as a contour integral
in a complex plane for integrands with multi-poles
in the plane. We show that 
the stable and efficient numerical integrations
along an appropriate contour for tensor integrals
as well as scalar ones appear in loop calculations of
the standard model.
This method is applicable, so far, up to two-loop/three-point
functions with arbitrary masses and arbitrary polynomials of
Feynman parameters in the numerator of the integrand. 
The basic idea of this method and some numerical results are
shown in this report.

\section{Feynman parameter representation of tensor integrals}
The tensor integral of a massless one-loop N-point graph with rank $M \leq N$ in a space-time
dimension of $n=4-2\eps$ can be written as,
\begin{eqnarray*}
T_{\underbrace{\mu \cdots \nu}_{M}}^{(N)}&=&\
\int  \frac{d^nk}{(2 \pi)^ni}
\frac{k_{\mu} \cdots k_{\nu}}{A_1 A_2\cdots A_N},
\end{eqnarray*}
where
\begin{eqnarray*}
A_i&=&(k+s_i)^2-m^2_i+i0,\\
s_i&=&\sum^i_{j=1}p_j,~~s_0=0, 
\end{eqnarray*}
and $p_i$ is a four momentum of an $i$'th external particle (incoming), 
$k_{\mu}$
a loop momentum, and $m_i$ the internal masses.
An infinitesimal imaginary part ($i0$) is included to obtain
analyticity of the integral $T_{\mu \cdots \nu}^{(N)}$.
The momentum integration can be done using Feynman's parameterization
which combines the propagators.
After the momentum integration, an ultra-violet pole is subtracted under some
renormalization scheme. 
Finally the tensor integral can be expressed using integrals of the type 
\begin{eqnarray}
I^{(N)}_{\underbrace{i \cdots k}_{M}}
&=&
\int [dx] 
\frac{x_i\cdots x_k}{\left(D_{N}-i0\right)^{N-2}},
\label{ti}
\end{eqnarray}
where
\begin{eqnarray*}
\int [dx]&=&\int^1_0 dx_1 \int^{1-x_1}dx_2 \cdots
\int^{1-\sum^{N-1}_{i=1} x_i} dx_{N-1}.
\end{eqnarray*}
An explicit form of the numerator,
and a relation between 
$T_{\mu \cdots \nu}^{(N)}$ and $I^{(N)}_{i \cdots k}$,
can be obtained straightforwardly depending on the diagram\cite{GRACEloop}.
The remaining task is to perform the parametric integration of 
$I^{(N)}_{i \cdots k}$ in Eq.(\ref{ti}),
which is called a {\it tensor integral} in this report.
(When the numerator of the integrand is unity, 
it is called a {\it scalar integral}.)

\section{one-loop/three-point}
A general formula of the tensor integral for the 
one-loop/three-point function with 
arbitrary masses is\cite{numeri1}
\begin{eqnarray*}
I^{(3)}&=&\int_0^1 dx \int_0^{1-x} dy \frac{f(x,y)}{D_{3}-i0}, \\
D_{3}&=&(M_1 x - M_2 y)^2-r x y+m_{31}^2 x +m_{32}^2 y+m_2^2,
\end{eqnarray*}
where
\begin{eqnarray*}
r&=&s-(M_1+M_2)^2, \\
m_{31}^2&=&-M_1^2+m_1^2-m_2^2, \\
m_{32}^2&=&-M_2^2+m_3^2-m_2^2, \\
M_1^2&=&p_1^2, \\
M_2^2&=&p_2^2, \\
s&=&(p_1+p_2)^2.
\end{eqnarray*}
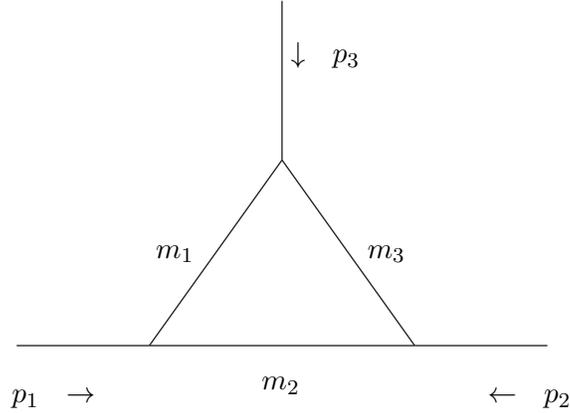
\begin{figure}[tb]
\begin{center}
\begin{picture}(300,200)(0,0)
\Line(150,180)(150,120)
\Line(50,50)(100,50)
\Line(250,50)(200,50)
\Line(150,120)(100,50)
\Line(100,50)(200,50)
\Line(200,50)(150,120)
\put(180,160){\makebox(0,0)[rc]{$\downarrow~~p_3$}}
\put(80,30){\makebox(0,0)[rc]{$p_1~~\rightarrow$}}
\put(260,30){\makebox(0,0)[rc]{$\leftarrow~~p_2$}}
\put(150,35){\makebox(0,0)[cc]{$m_2$}}
\put(110,85){\makebox(0,0)[cc]{$m_1$}}
\put(190,85){\makebox(0,0)[cc]{$m_3$}}
\end{picture}
\caption{1-loop/3-point diagram}
\end{center}
\label{1l3p}
\end{figure}
A numerator of the integrand, $f(x,y)$, can be any polynomial of
Feynman parameters $x$ and $y$ with rank $M\leq3$.
Momentum and mass assignments are shown in Figure~1.

The integration region is a 2-dimensional simplex with three sides.
Those sides are 
\begin{eqnarray*}
L_x&=&\{(x,y)|0\le x \le 1, y=0\}, \\
L_y&=&\{(x,y)|x=0, 0\le y \le 1\}, \\ 
L_{xy}&=&\{(x,y)|y=1-x,0\le x \le 1\}.
\end{eqnarray*}
Singular points (second pole) of the integrand form a
parabola (or hyperbola), which is 
\begin{eqnarray*}
{\cal P}&=&\{(x,y)|D_3=0\}.
\end{eqnarray*}
The focal point of the parabola is on the opposite side of the origin
with respect to the parabola line. 
When one integration-variable is fixed, the remaining one-parameter 
integration
may intersect a pole on the line ${\cal P}$. This integration can be done
numerically as a contour integral along an appropriate contour with correct
analyticity. For a stable integration, it must be avoided that
the contour intersects more than one pole, or that a pole is at the
end point of the contour. These requirements are satisfied
by taking an appropriate coordinate system instead of a simple $(x,y)$ 
coordinate of Feynman parameters. 
Basically we take polar-coordinates with an origin on the
side $L_{xy}$.
Coordinate systems we used are
categorized by the following cases;
\begin{enumerate}
\item ${\cal N}[(L_x \cup L_y \cup L_{xy})\cap{\cal P}]=0$: \\
${\cal N}[{\cal A}]$ are number of elements of a set ${\cal A}$.
There is no singular points in the integration
region. The origin is set at the nearest point to the line ${\cal P}$
on the side $L_{xy}$. 
\item ${\cal N}[L_{xy} \cap {\cal P}]=1$ $\wedge$
${\cal N}[(L_x \cup L_y)\cap{\cal P}]=0$: \\
The origin of the coordinate system is set at a point such as
$\{L_{xy} \cap {\cal P}\}$. Though a pole is at the origin,
the singularity is
canceled out against the Jacobian on the numerator, i.e. the measure
at the origin is zero in the polar-coordinates. Then the 
integration is free from singularities.
\item ${\cal N}[L_{xy} \cap {\cal P}]=2$ $\wedge$ 
${\cal N}[(L_x \cup L_y)\cap{\cal P}]=0$: \\
The origin is set at the center of two elements of 
$\{L_{xy} \cap {\cal P}\}$. 
Then the contour along $r$ intersects the line ${\cal P}$ only
once for any value of $\phi$. We can take an appropriate contour
avoiding the pole with correct analyticity.
%
\item ${\cal N}[L_{xy} \cap {\cal P}]=2$ $\wedge$ 
${\cal N}[L_x\cap{\cal P}]=2$ $\wedge$
${\cal N}[L_y\cap{\cal P}]=2$: \\
The integration region is divided into three regions by two lines,
one connects between centers of two elements of $L_{xy} \cap {\cal P}$
and those of $L_x\cap{\cal P}$, the other connects between
centers of two elements of $L_{xy} \cap {\cal P}$ and $L_y\cap{\cal P}$.
The origins of the coordinate frames are set at 
the three corner of the simplex,
$(0,0)$, $(1,0)$ and $(0,1)$. In each region, 
the contour along $r$ intersects the line ${\cal P}$ only
once for any value of $\phi$. We can take an appropriate contour
avoiding the pole with correct analyticity.
\end{enumerate}

Numerical integrations are performed based on the 
`Good Lattice Point (GLP) Method\cite{glp}',
which uses a deterministic series of numbers.
For smooth functions 
it shows very efficient convergence compared with a Monte Carlo
integration based on random sampling. Moreover both imaginary
and real parts can be integrated simultaneously.

Here we show several examples of one-loop/three-point functions.
\newpage
\begin{itemize}
\item Case~1:$m_1=m_3=M_1=M_2=150$ GeV, $m_2=91$ GeV, $f(x,y)=1$.
\begin{center}
{\bf Table~1}
\begin{tabular}{|r|c||r||r|r|r|}
  \hline
  $\sqrt{s}$ & real/imag. & analytic\cite{suppl}~~~ & NCI result ~~~~ & error~~~~~~&calls \\
  \hline
  \hline
310 & real & $0.1110 \times 10^{-3}$ &
             $0.110975\times 10^{-3}$&$0.23\times10^{-10}$ & 350 \\
~   & imag.& $0.7162 \times 10^{-4}$ & 
             $0.716174\times 10^{-4}$&~ & ~  \\
\hline
500 & real & $0.2285 \times 10^{-5}$ &
             $0.228498\times 10^{-5}$&$0.39\times10^{-10}$ & 350 \\
~   & imag.& $0.4731 \times 10^{-4}$ & 
             $0.473072\times 10^{-4}$&~ & ~  \\
\hline
1000 & real & $-0.6103 \times 10^{-5}$ &
              $-0.610293\times 10^{-5}$&$0.22\times10^{-11}$ & 582 \\
~    & imag.& $0.1551 \times 10^{-4}$ & 
              $0.155066\times 10^{-4}$&~ & ~  \\
\hline
\end{tabular}
\end{center}
This is an infrared-divergence free integral
with only heavy particles involved.
Very good agreements can be obtained between analytical results
(formulae given in ref.\cite{suppl}) and the NCI method
as shown in the above table with only several hundred sampling points. 
\item Case~2:$m_1=m_3=M_1=M_2=m_e$, $\sqrt{s}=1000$ GeV, $f(x,y)=1$.
\begin{center}
{\bf Table~2}
  \begin{tabular}{|r|c||r||r|r|r|}
  \hline
  $m_2 (GeV)$ & real/imag. & analytic\cite{suppl}~~~~ & NCI results~~~~ & error~~~~~~&calls \\
  \hline
  \hline
$10^{-5}$ & real & 
             $-0.641175 \times 10^{-3}$ &
             $-0.641175 \times 10^{-3}$&$0.20\times10^{-9}$ & 6732 \\
~   & imag.& $~0.115741 \times 10^{-3}$ & 
             $~0.115740 \times 10^{-3}$&~ & ~  \\
\hline
$10^{-7}$ & real & 
             $-0.907973 \times 10^{-3}$ &
             $-0.907973 \times 10^{-3}$&$0.22\times10^{-7}$ & 6732 \\
~   & imag.& $~0.144676 \times 10^{-3}$ & 
             $~0.144674 \times 10^{-3}$&~ & ~  \\
\hline
$10^{-9}$ & real & 
              $-0.117483 \times 10^{-2}$ &
              $-0.117479 \times 10^{-2}$&$0.10\times10^{-6}$ & 28621 \\
~    & imag.& $~0.173611 \times 10^{-3}$ & 
              $~0.173618 \times 10^{-3}$&~ & ~  \\
\hline
\end{tabular}
\end{center}
This is an infrared-divergent case with a fictitious photon mass of 
$10^{-5}$ to $10^{-9}$ GeV.
Very good agreements can be obtained between analytical results
and the NCI, even for the infrared-divergent case with
several thousand sampling points. 
\item Case~3:$m_1=m_3=80$ GeV, $M_1=M_2=170$ GeV, $M_2=5$ GeV, $\sqrt{s}=357$ GeV \\
\begin{center}
{\bf Table~3}
  \begin{tabular}{|c|c||r||r|r|r|}
  \hline
  $f(x,y)$ & real/imag. & $FF\cite{ff}$~~~~~~~~~ & NCI results~~~~ & error~~~~~~&calls \\
  \hline
  \hline
$1$ & real & 
             $-0.623169 \times 10^{-4}$ &
             $-0.623169 \times 10^{-4}$&$0.18\times10^{-14}$ & 1567 \\
~   & imag.& $~0.132340 \times 10^{-4}$ & 
             $~0.132340 \times 10^{-4}$&~ & ~  \\
  \hline
$x$ & real & 
             $-0.171520 \times 10^{-4}$ &
             $-0.171520 \times 10^{-4}$&$0.43\times10^{-9}$ & 1567 \\
~   & imag.& $~0.579780 \times 10^{-5}$ & 
             $~0.579780 \times 10^{-5}$&~ & ~  \\
  \hline
$x^2$ & real & 
             $-0.859550 \times 10^{-5}$ &
             $-0.859550 \times 10^{-5}$&$0.13\times10^{-9}$ & 1567 \\
~   & imag.& $~0.487062 \times 10^{-5}$ & 
             $~0.487063 \times 10^{-5}$&~ & ~  \\
  \hline
$x\times y$ & real & 
             $-0.324021 \times 10^{-5}$ &
             $-0.324021 \times 10^{-5}$&$0.11\times10^{-9}$ & 1567 \\
~   & imag.& $~0.296541 \times 10^{-6}$ & 
             $~0.296540 \times 10^{-6}$&~ & ~  \\
  \hline
\end{tabular}
\end{center}
\end{itemize}
The tensor integration can be done as well as the scalar one as shown
in Table~3. NCI results show a very good agreement with those from the $FF$ 
package\cite{ff} developed by van Oldenborgh.

\section{one-loop/four-point}
A general formula for the one-loop/four-point function 
with arbitrary masses is\cite{numeri3}
\begin{eqnarray*}
I^{(4)}&=&\int_0^1 dz \int_0^{1-z} dy \int_0^{1-y-z} 
dx \frac{f(x,y,z)}{D_4-i0}, \\
D_{4}&=&(M_1 x - M_2 y)^2-r x y+m_{41}^2 x +m_{42}^2 y+m_{43}^2,
\end{eqnarray*}
where
\begin{eqnarray*}
t&=&(p_2+p_3)^2, \\
u&=&(p_1+p_3)^2, \\
m_{41}^2&=&-M_1^2+m_1^2-m_2^2-z(s+u-2M_1^2-M_2^2-M_3^2), \\
m_{42}^2&=&-M_2^2+m_3^2-z(t+M_2^2-M_3^2), \\
m_{43}^2&=&m_2^2+(-M_3^2+m_4^2-m_2^2)z+tz^2
\end{eqnarray*}
A numerator of the integrand, $f(x,y,z)$, can be any polynomial of
Feynman parameters, $x$, $y$ and $z$ with rank $M\leq4$.
Momentum and mass assignments are shown in Figure~2.
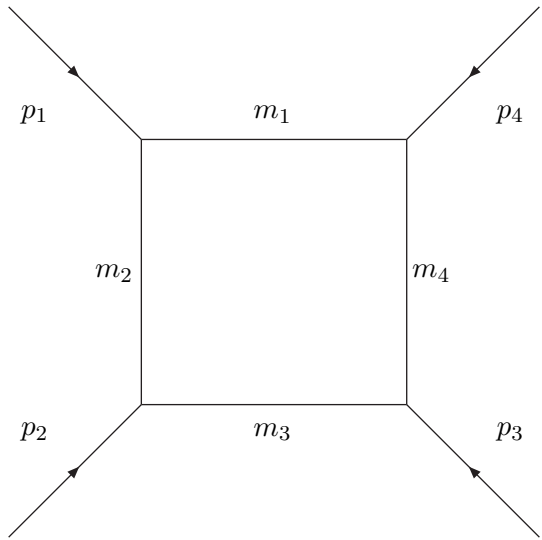
\begin{figure}[tb]
\begin{center}
\begin{picture}(300,200)(0,0)
\ArrowLine( 50,  0)(100, 50)
\ArrowLine(250,  0)(200, 50)
\ArrowLine(250,200)(200,150)
\ArrowLine( 50,200)(100,150)
\Line(100,150)(100, 50)
\Line(100, 50)(200, 50)
\Line(200, 50)(200,150)
\Line(200,150)(100,150)
\put(150,160){\makebox(0,0)[cc]{$m_1$}}
\put( 90,100){\makebox(0,0)[cc]{$m_2$}}
\put(150, 40){\makebox(0,0)[cc]{$m_3$}}
\put(210,100){\makebox(0,0)[cc]{$m_4$}}
\put( 60,160){\makebox(0,0)[cc]{$p_1$}}
\put( 60, 40){\makebox(0,0)[cc]{$p_2$}}
\put(240, 40){\makebox(0,0)[cc]{$p_3$}}
\put(240,160){\makebox(0,0)[cc]{$p_4$}}
\end{picture}
\caption{1-loop/4-point diagram}
\end{center}
\label{1l4p}
\end{figure}
The integration region is a 3-dimensional simplex.
In order to perform the contour integral very efficiently
in this three-dimensional integration-parameter space,
we have introduced a new coordinate system as shown in
Figure~3, named `Wedge coordinate system'.
\begin{figure}[tbh]
\begin{center}
\includegraphics[width=0.8\linewidth]{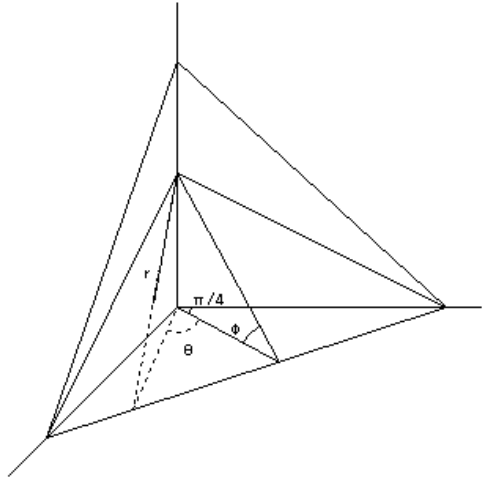}
\caption{
Parametrization of the wedge coordinate
}
\label{wedge}
\end{center}
\end{figure}
A relation between an usual Cartesian coordinate system and the 
wedge coordinate system is given as
\begin{eqnarray*}
x&=&r\cos{\phi'} \sin{\left(\theta+\frac{\pi}{4}\right)}, \\
y&=&r\cos{\phi'} \cos{\left(\theta+\frac{\pi}{4}\right)}, \\
z&=&(r_{max}-r)\sin{\phi'},
\end{eqnarray*}
where
\begin{eqnarray*}
\tan{\phi'}&=&\cos{\theta} \tan{\phi}, \\
r_{max}&=&\sqrt{\frac{\tan^2{\phi}}{2}+\frac{1}{2 \cos^2{\theta}}}.
\end{eqnarray*}
The Feynman parameter integration now becomes
\begin{eqnarray*}
I^{(4)}
&=&\int_0^1 dz \int_0^{1-z} dy \int_0^{1-y} dx \frac{f(x,y,z)}{D_4-i0} \\
~&=&\int_0^{r_{max}} dr \int_{-\pi/4}^{\pi/4} d\theta 
\int_0^{\tan^{-1}{\sqrt{2}}} d\phi~~r(r_{max}-r)\frac{\cos{\theta}\cos^3{\phi'}}
{\cos^2{\phi}} \frac{f(r,\theta,\phi)}{D_4-i0}.
\end{eqnarray*}
When $\phi$ is fixed, the remaining integration region is a triangle.
The intersection between this triangle and a hyper-surface which satisfies
$D_4=0$ is a parabola (or hyperbola) similar to the one-loop/three-point
case. The variable $r$ is extended into the complex plane.
When both $\phi$  and $\theta$ are fixed, 
we can take an appropriate contour for $r$ which crosses a pole only once
for any $\phi$  and $\theta$.

Here we show several examples of one-loop/four-point functions.
\begin{itemize}
\item Case~1:$e^+e^- \rightarrow W^+ W^-$. \\
$m_1=m_3=80$ GeV, $m_2=0$, $m_4=91$ GeV, $M_1=M_2=m_e$, $M_3=M_4=80$ GeV,
$\sqrt{s}=500$ GeV, $\theta=\angle(P_1,P_4)$.
\begin{center}
{\bf Table~4}
\begin{tabular}{|r|c||r||r|r|r|}
  \hline
$\cos{\theta}$
& real/imag. & $FF\cite{ff}$~~~~~~~~~& NCI results~~~~ & error~~~~~~&calls \\
  \hline
  \hline
$-0.5$ & real & 
$-0.459774 \times 10^{-9}$ & 
$-0.459774 \times 10^{-9}$ & $0.67\times 10^{-13}$ & 1567 \\ 
$ ~  $ & imag. & 
$ 0.711583 \times 10^{-9}$ & 
$ 0.711582 \times 10^{-9}$ & ~&~ \\ 
  \hline
$ 0.0$ & real & 
$-0.588538 \times 10^{-9}$ & 
$-0.588539 \times 10^{-9}$ & $0.10\times 10^{-12}$ & 1567 \\ 
$ ~  $ & imag. & 
$ 0.962985 \times 10^{-9}$ & 
$ 0.962985 \times 10^{-9}$ & ~&~ \\ 
  \hline
$ 0.5$ & real & 
$-0.849170 \times 10^{-9}$ & 
$-0.849167 \times 10^{-9}$ & $0.10\times 10^{-13}$ & 2553 \\ 
$ ~  $ & imag. & 
$ 0.155695 \times 10^{-8}$ & 
$ 0.155695 \times 10^{-8}$ & ~&~ \\ 
  \hline
\end{tabular}
\end{center}
Though a massless particle (neutrino) appears in the loop in this case,
this is infrared-divergence free. The NCI with several thousand sampling 
points gives very good agreement with $FF$.
\item Case~2:$e^+e^- \rightarrow z z$. \\
$m_1=m_3=m_4=m_e$, $m_2=91$ GeV, $M_1=M_2=m_e$, $M_3=M_4=91$ GeV,
$\sqrt{s}=500$ GeV, $\theta=\angle(P_1,P_4)$.
\begin{center}
{\bf Table~5}
\begin{tabular}{|r|c||r||r|r|r|}
  \hline
$\cos{\theta}$
& real/imag. & $FF\cite{ff}$~~~~~~~~~& NCI results~~~~ & error~~~~~~&calls \\
  \hline
  \hline
$-0.5$ & real & 
$-0.89532 \times 10^{-9}$ & 
$-0.89540 \times 10^{-9}$ & $0.54\times 10^{-12}$ & $2M$ \\
$ ~  $ & imag. & 
$-0.23060 \times 10^{-9}$ & 
$-0.23049 \times 10^{-9}$ & ~&~ \\ 
  \hline
$ 0.0$ & real & 
$-0.11548 \times 10^{-8}$ & 
$-0.11549 \times 10^{-8}$ & $0.78\times 10^{-11}$ & $2M$ \\ 
$ ~  $ & imag. & 
$-0.33447 \times 10^{-9}$ & 
$-0.33433 \times 10^{-9}$ & ~&~ \\ 
  \hline
$ 0.5$ & real & 
$-0.17129 \times 10^{-8}$ & 
$-0.17134 \times 10^{-8}$ & $0.96\times 10^{-11}$ & $2M$ \\ 
$ ~  $ & imag. & 
$ 0.60873 \times 10^{-9}$ & 
$ 0.60862 \times 10^{-9}$ & ~&~ \\ 
  \hline
\end{tabular}
\end{center}
A light particle (an electron) exists in the loop in this case. 
Although it requires high statistics of about two million sampling points
even by the GLP, the result shows good agreement with $FF$.
\end{itemize}
Though only scalar integrations are shown here, tensor integrations
with an arbitrary polynomial in the numerator can be done as well.

\section{two-loop/two-point}
A general formula of the two-loop/two-point 
function with arbitrary masses is\cite{numeri4}
\begin{eqnarray*}
I^{(2)}&=&\int_0^1 dx_1 dx_2 dx_3 dx_4 dx_5 
\delta\left(1-\sum_{i=1}^5 x_i\right)\frac{f(x_i)}{DC-i0},
\end{eqnarray*}
where
\begin{eqnarray*}
D&=&-p^2\left(x_5\left(x_1+x_3\right)\left(x_2+x_4\right)
+\left(x_1+x_2\right)x_3 x_4+\left(x_3+x_4\right)x_1 x_2\right)+C{\bar M}^2,\\
C&=&\left(x_1+x_2+x_3+x_4\right)x_5+\left(x_1+x_2\right)\left(x_3+x_4\right),\\
{\bar M}^2&=&\sum_{i=1}^5 x_i m_i^2,
\end{eqnarray*}
and $m_i$ are the internal masses. A particle number assignment 
is shown in Figure~4. Here $f(x_i)$ is any polynomial of Feynman parameters
$x_i$ with rank $M\leq2$.
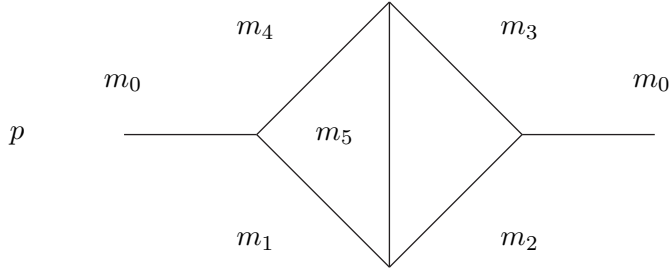
\begin{figure}[tb]
\begin{center}
\begin{picture}(600,200)(0,0)
\Line(100,100)(150,100)
\Line(150,100)(200, 50)
\Line(200, 50)(250,100)
\Line(250,100)(200,150)
\Line(200,150)(150,100)
\Line(200,150)(200, 50)
\Line(250,100)(300,100)
\put( 60,100){\makebox(0,0)[cc]{$p$}}
\put(100,120){\makebox(0,0)[cc]{$m_0$}}
\put(300,120){\makebox(0,0)[cc]{$m_0$}}
\put(150,140){\makebox(0,0)[cc]{$m_4$}}
\put(150, 60){\makebox(0,0)[cc]{$m_1$}}
\put(250,140){\makebox(0,0)[cc]{$m_3$}}
\put(250, 60){\makebox(0,0)[cc]{$m_2$}}
\put(180,100){\makebox(0,0)[cc]{$m_5$}}
\end{picture}
\label{2l2p}
\caption{2-loop/2-point diagram}
\end{center}
\end{figure}

Feynman parameter integration with four independent variables
must be performed after $x_5$ integration eliminates a delta-function.
It is very hard to make any intuitive optimization
for the integration because four-dimensional space is beyond our 
imagination. 
We simply employ the following coordinate system:
At first one-point on the edge-simplex of 
$x_1+x_2+x_3+x_4=1$ is chosen.
Then take a variable $r$ as a distance between a origin of the
coordinate ($x_1=x_2=x_3=x_4=0$) and the point on a
straight line from the origin to the point in the edge-simplex.
This $r$ is extended into the complex plane and chosen to follow
an appropriate contour to avoid singularities.
Here we show two examples of two-loop/two-point functions.
\begin{itemize}
\item Case~1:$m_1=m_2=m_3=m_4=m_0=150$ GeV and
$m_5=91.17$ GeV.
\begin{center}
{\bf Table~6}\\
\begin{tabular}{|r|c||r||r|}
  \hline
$p^2/m_0^2$ & real/imag. & Kreimer\cite{mainz} & NCI results \\ 
\hline \hline
2.0& real & 2.664 & 2.6647(2) \\ \hline
4.0& real & 25.85 & 25.87(2)  \\ \hline
4.1& real & 21.506&21.47(3) \\
~  & imag.& 12.16 &12.15(3) \\ \hline
4.3& real & 15.246&15.19(3) \\
~  & imag.& 17.013&17.29(3) \\ \hline
5.0& real & 3.668 &3.63(4) \\
~  & imag.& 19.503&19.48(4) \\ \hline
7.0& real & -6.7939&-6.78(4)  \\
~  & imag.& 13.918 & 13.89(4) \\ \hline
10.0& real & -8.793 & -8.792(1)  \\
~  & imag.& 8.865 & 8.846(1) \\ \hline
\end{tabular}
\end{center}
In this case, integrations are done by Monte Carlo method
using BASES\cite{bases}. NCI results show good agreement with the 
analytical results by Kreimer\cite{mainz}, within
the statistical error of the Monte Carlo integration.
\item Case~2:$m_1=1$ GeV, $m_2=2$ GeV, $m_3=4$ GeV, $m_4=5$ GeV,
$m_5=3$ GeV.
\begin{center}
{\bf Table~7}
\begin{tabular}{|r|c||r||r|}
  \hline
$p^2(GeV^2)$ & real/imag. & Bauberger et al.\cite{bohm} & NCI results \\ 
\hline \hline
0.1& real & -0.287238(3) & -0.287240(6) \\ \hline
0.5& real & -0.294592(3) & -0.294594(6) \\ \hline
1.0& real & -0.304521(3) & -0.304523(6) \\ \hline
5.0& real & -0.452520(3) & -0.452527(9) \\ \hline
10.0& real & -0.488153(2) & -0.48807(5) \\
~  & imag. & -0.353217(2) & -0.35307(4) \\ \hline
50.0& real & 0.173901(2)  &  0.17397(2)    \\
~  & imag. & -0.118080(2) & -0.11805(2) \\ \hline
\end{tabular}
\end{center}
\end{itemize}
This is an example with a very simple mass assignment. The result of the NCI 
method is compared with that from the analytical calculation obtained by
Bauberger and B{\" o}hm\cite{bohm}, and gives also very good agreement. 
The same example is treated using the SBT relation by Passarino 
and Uccirati\cite{passarino2} with very good agreement too.

Though only scalar integrations are shown here, tensor integrations
with an arbitrary polynomial in the numerator can be done as well.

\section{two-loop/three-point}
\begin{figure}[tb]
\begin{center}
\begin{picture}(187,201) (248,-49)
\SetWidth{0.5}
\SetColor{Black}
\ArrowLine(345,137)(345,81)
\Line(287,-25)(345,81)
\Line(389,-23)(345,81)
\ArrowLine(389,-23)(376,7)
\Line(324,42)(376,7)
\Line(303,4)(361,44)
\ArrowLine(287,-25)(303,4)
\Text(367,63)[lb]{\Large{\Black{$m_2$}}}
\Text(310,-13)[lb]{\Large{\Black{$m_3$}}}
\Text(351,-13)[lb]{\Large{\Black{$m_6$}}}
\Text(267,22)[lb]{\Large{\Black{$m_4$}}}
\Text(387,22)[lb]{\Large{\Black{$m_5$}}}
\Text(357,136)[lb]{\Large{\Black{$p_3$}}}
\Text(248,-45)[lb]{\Large{\Black{$p_1$}}}
\Text(405,-49)[lb]{\Large{\Black{$p_2$}}}
\Text(287,63)[lb]{\Large{\Black{$m_1$}}}
\end{picture}
\caption{
2-loop vertex diagram (non-planer type)
}
\label{2l3p}
\end{center}
\end{figure}
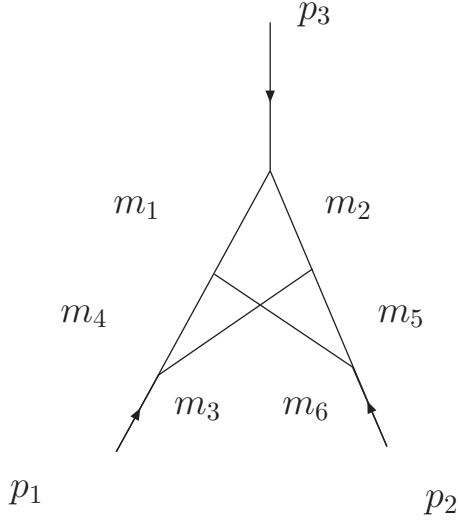
A general formula of the two-loop/three-point 
function (non-planar configuration) 
with arbitrary masses 
can be expressed as
\begin{eqnarray*}
I^{(3)}&=&\frac{1}{8}\int_0^1dz_1 dz_2 dz_3 \delta\left(1-\sigma z\right)
z_1 z_2 z_3 \int_{-1}^{1}dy_1 dy_2 dy_3 \frac{f(y_i,z_j)}
{\left(D_{3}-i0\right)^2},
\end{eqnarray*}
after some transformation of the original Feynman parametrization\cite{2loop2}.
Here the denominator is
\begin{eqnarray*}
D_{3}&=&^t{\overrightarrow y}A{\overrightarrow y}+^t{\overrightarrow b}
\cdot{\overrightarrow y}+c,
\end{eqnarray*}
where
\begin{eqnarray*}
{\overrightarrow y}&=&
\left(\begin{array}{c}y_1\\y_2\\y_3\end{array}
\right), \\
\end{eqnarray*}
\begin{eqnarray*}
A&=&\frac{1}{4}\left(
\begin{array}{ccc}
-z_1^2(z_2+z_3)s_1&z_1 z_2 z_3(-s_1-s_2+s_3)/2&z_1 z_2 z_3(-s_1+s_2-s_3)/2 \\
z_1 z_2 z_3(-s_1-s_2+s_3)/2&-z_2^2(z_3+z_1)s_2&z_1 z_2 z_3(+s_1-s_2-s_3)/2 \\
z_1 z_2 z_3(-s_1+s_2-s_3)/2&z_1 z_2 z_3(+s_1-s_2-s_3)/2&-z_3^2(z_1+z_2)s_3 
\end{array}
\right),
\end{eqnarray*}
\begin{eqnarray*}
{\overrightarrow b}&=&\frac{1}{2}U
\left(
\begin{array}{c}
z_1(m_3^2-m_4^2) \\
z_2(m_5^2-m_6^2) \\
z_3(m_2^2-m_1^2)
\end{array}
\right),
\end{eqnarray*}
\begin{eqnarray*}
c&=&\frac{1}{4}U\left[z_1 s_1+z_2 s_2+z_3 s_3
-2(m_3^2+m_4^2)z_1
-2(m_5^2+m_6^2)z_2
-2(m_1^2+m_2^2)z_3
\right], \\
U&=&z_1 z_2+z_2 z_3+z_3 z_1,
\end{eqnarray*}
and $s_i=p_i^2$, and $m_i$ are the internal masses.
Here $^t{\overrightarrow v}$ gives a transposed vector of ${\overrightarrow v}$.
For the $y$ integration, we employ the SBT relation as
\begin{eqnarray*}
\frac{1}{D_3^2}&=&\frac{1}{\beta}\left[1
+\frac{1}{2}~^t\left({\overrightarrow y}-{\overrightarrow \eta}\right)
\cdot {\overrightarrow \partial_y}
\right]\frac{1}{D_3}, \\
\beta&=&c-\left(^t{\overrightarrow b}A^{-1}{\overrightarrow b}\right), \\
{\overrightarrow \eta}&=&-A^{-1}{\overrightarrow b} \\
{\overrightarrow \partial_y}&=&
\left(
\frac{\partial~~}{\partial y_1},
\frac{\partial~~}{\partial y_2},
\frac{\partial~~}{\partial y_3}
\right).
\end{eqnarray*}
The $y_i$ are left as real variables. In order to avoid singularities
in $\beta^{-1}$ and $D_3^{-1}$, the $z_i$ are extended into the complex plane.
After the $z_3$ integration to eliminate the $\delta$-function,
the polar coordinates, $z_1=r_z\sin \theta_z,~z_2=r_z\cos \theta_z$ are
introduced. 
This $r_z$ is extended into the complex plane and chosen 
along an appropriate contour to avoid singularities.
Here we show one example of a two-loop/three-point function.
\begin{itemize}
\item $m_1=m_2=m_4=m_5=150$ GeV,
$m_3=m_6=91.17$ GeV, and $s1=s2=150^2~{\rm GeV}^2$.
\begin{center}
{\bf Table~8} \\
\begin{tabular}{|r||r||r|}
  \hline
$s_3/m_1^2$ &Fujimoto et al.\cite{2loop2}&  NCI results~~~~~ \\ 
\hline \hline
4.5&$2.09(2)\times10^{-9}$&$2.049(2)\times10^{-9}$ \\ \hline
5&$1.43(1)\times10^{-9}$& $1.464(2)\times10^{-9}$ \\ \hline
12&$-8.37(3)\times10^{-10}$&$-8.24(4)\times10^{-10}$ \\ \hline
20&$-5.61(3)\times10^{-10}$&$-5.60(1)\times10^{-10}$ \\ \hline
\end{tabular}
\end{center}
\end{itemize}
The Monte Carlo integration package, BASES, is used for the numerical 
integration. The results agree well with those in 
refs.\cite{2loop2,kreimer} 

Though only scalar integrations are shown here, tensor integrations
with arbitrary polynomial in the numerator can be done as well.

\section{Numerical evaluations of the Hypergeometric functions}
So far we discussed in previous sections
loop-integrals in the standard model with
arbitrary masses. In this section we would like to
treat a massless theory such as QCD. A general one-loop/four-point function
in a massless theory can be expressed by the
following tensor integrals:
\begin{eqnarray*}
I^{(4)}(s,t;n_x,n_y,n_z)&=&\left(4 \pi \mu^2\right)^{-\eir}
\int_0^1 dx~dy~dz
\frac{x^{n_x}y^{n_y}z^{n_z}}{(D_4-i0)^{2-\eir}},\\
D_4&=&-xzs-y(1-x-y-z)t,
\end{eqnarray*}
where $s=(p_1+p_2)^2$, and $t=(p_1+p_4)^2$.
Here we set all particles massless.
In order to cure an infrared divergence, the space-time dimension
is set to be $n=4+2\eir$ after the $\overline{MS}$ renormalization.
The tensor integration can be done analytically and be represented
by a finite number of terms with Beta and
Hypergeometric functions as\cite{kurihara};
\begin{eqnarray*}
&~&I^{(4)}(s,t;n_x,n_y,n_z)
=\frac{1}{s~t}\frac{B(n_x+\eir,n_y+n_z+\eir)}{1-\eir} \nonumber \\
&\times&\biggl[
\left(\frac{-\t}{4 \pi \mu^2}\right)^{\eir}
\left(\frac{-t}{s}\right)^{n_x}
\frac{n_x}{\prod_{j=1}^{n_x}({n_x}-j+\eir)}B(1+n_z,n_x+n_y+\eir)
\nonumber \\
&\times&\hyg\left(1+n_x,n_x+n_z+\eir,1+n_x+n_y+n_z+\eir,-\frac{\u}{\s}\right)
\nonumber \\
&+&\left(\frac{-\s}{4 \pi \mu^2}\right)^{\eir}
\sum_{l=0}^{n_x} \left(\frac{-s}{t}\right)^l
\frac{\prod_{j=1}^l(l-j-n_x)}{\prod_{j=1}^l(l-j+\eir)}B(1+n_y,l+n_z+\eir)
\nonumber \\
&\times&\hyg\left(1+l,l+n_z+\eir,1+l+n_y+n_z+\eir,-\frac{\u}{\bar t}\right)
\biggr], \\
{\s}&=&s+i0, \nonumber \\
{\t}&=&t+i0,  \nonumber\\
{\bar t}&=&t-i0, \nonumber \\
{\u}&=&u+i0=(p_1+p_3)^2+i0.\nonumber
\end{eqnarray*}
For the infrared-finite case, evaluation of the loop integral
can be done if we can perform 
the numerical calculation of the Hypergeometric function
of a type
\begin{eqnarray*}
\hyg\left(1+l,l+m,1+l+m+n,z\right)&=&\frac{\Gamma(1+l+m+n)}{\Gamma(l+m)
\Gamma(1+n)}\int_0^1d\tau\frac{\tau^{l+m-1}(1-\tau)^n}{(1-z\tau)^{1+l}}
\end{eqnarray*}
where $l,m,n>0$ are integer numbers and $z$ a complex variable.
This integration can be performed easily using the NCI. 

\begin{center}
\begin{tabular}{|ccc|c||r||r|c|}
  \hline
$l$ & $m$ & $n$ &real/imag.&$Mathematica\cite{mathematica}$~~~~& NCI~~~~~~~~~~~~~~&calls  \\ \hline \hline
1&1&1& real& $-0.1453322029\times10^{-1}$ & $-0.1453322029\times10^{-1}$ &4149\\
~&~&~& imag.&$-0.1507964474\times10^{0~~}$& $-0.1507964474\times10^{0~~}$ &\\ \hline
1&2&3& real& $0.8417767168\times10^{-1}$ & $0.8417767169\times10^{-1}$ & 4149\\
~&~&~& imag.&$-0.2290221044\times10^{0~~}$ & $-0.2290221044\times10^{0~~}$ &\\ \hline
2&1&1& real& $0.1087664688\times10^{-1}$ & $0.1087664688\times10^{-1}$ &4149\\
~&~&~& imag.&$0.2638937829\times10^{-1}$ & $0.2638937829\times10^{-1}$ &\\ \hline
2&3&4& real& $-0.2890568082\times10^{-1}$ & $-0.2890568082\times10^{-1}$ &6732\\
~&~&~& imag.&$0.5578978464\times10^{-1}$ & $0.5578978464\times10^{-1}$ &\\ \hline
3&1&2& real& $-0.1026721798\times10^{-1}$ & $-0.1026721798\times10^{-1}$ &6732\\
~&~&~& imag.&$-0.5654866776\times10^{-2}$ & $-0.5654866776\times10^{-2}$ &\\ \hline
3&4&5& real& $0.8121358824\times10^{-2}$ & $0.8121358824\times10^{-2}$ &6732\\
~&~&~& imag.&$-0.1274636264\times10^{-1}$ & $-0.1274636264\times10^{-1}$ &\\ \hline
\end{tabular}
{\bf Table~9} 
{\footnotesize numerical calculation of the Hypergeometric function at $z=10$.}
\end{center}
The results from the NCI method with the GLP numerical integration at
$z=10+i0$ are shown in Table~9 comparing them with those obtained
by $Mathematica$\cite{mathematica}.
The ten-digit agreement with $Mathematica$ can be obtained 
with only four to seven thousand sampling points in the GLP.

\section{Summary}
We have developed a fully numerical method,
named `numerical contour integration (NCI)' method, 
to calculate loop integrals.
Loop integrals can be interpreted as a contour integral
in a complex plane for an integrand with multi-poles
in the plane. 
For the one-loop/three-point case with arbitrary masses in the standard model,
both scalar and tensor integrals have been obtained by the NCI method with
five to seven digits accuracy.
For the one-loop/four point case, scalar integrals 
have been calculated with about 0.1\% accuracy. For the above two cases,
the `Good Lattice Point' method is used for the numerical integration.
Those results are compared with analytical calculation or $FF$ and show
a very good agreement. The NCI method has been applied also to 
two-loop integrals. For two- and three-point integrals at the two-loop
level, it is demonstrated that the NCI method can give numerical results
of the loop integrals with good accuracy using the Monte Carlo
integration package BASES. Those results show a good agreement with
previous calculations. The extension to the general tensor cases
is straight forward, since the NCI is a purely numerical method.
Moreover it is shown that 
the numerical evaluation of the Hypergeometric function, which 
appears in the one-loop/four-point tensor-integrals in massless QCD,
has been performed with ten-digit accuracy using the GLP method.

\vskip 1cm
Authors would like to thank Drs. J.~Fujimoto, T.~Ishikawa and Y.~Shimizu
for continuous discussions
on this subject and their useful suggestions.
We are grateful to 
Prof. J.~Vermaseren for his proofreading the manuscript 
which improved the English very much.

This work was supported in part by the Ministry of Education, Science
and Culture under the Grant-in-Aid No. 11206203 and 14340081.

\end{document}